\documentclass[showpacs,twocolumn,amsmath,amssymb]{revtex4-1}

\usepackage{amsmath} % need for subequations

    \usepackage{graphicx}
    \usepackage{color}
    \usepackage{dcolumn}
    \usepackage{bm}
    \usepackage{xspace}
    \usepackage{hyperref}

\usepackage{calc}

   \newcommand{\Ignore}[1]{}%_YES_this_DOES_ignore_text
\usepackage[usenames,dvipsnames]{xcolor}
\usepackage{ esint }
\usepackage{ amssymb }
\usepackage{graphicx}  
\usepackage{fancybox}  

\newcommand{\figref}[1]{\ref{#1}}

\newcommand{\nd}[1]{$#1$}%nature_dollar_for_introduction_of_variables

\newcommand{\new}[1]{\textcolor{blue}{#1}}%switches_to_NO_color
\renewcommand{\new}[1]{#1}%switches_to_NO_color
\begin{document}

\title{\new{Wigner flow reveals topological order in quantum phase space  dynamics}}

\author{Ole Steuernagel, Dimitris Kakofengitis and Georg Ritter}

\affiliation{School of Physics, Astronomy and Mathematics, University of
Hertfordshire, Hatfield, AL10 9AB, UK}
\email{O.Steuernagel@herts.ac.uk}

\date{\today}

\begin{abstract}
  The behaviour of classical mechanical systems is characterised by
  their phase portraits, the collections of their trajectories.
  Heisenberg's uncertainty principle precludes the existence of
  sharply defined trajectories, which is why traditionally only the
  time evolution of wave functions is studied in quantum
  dynamics. These studies are quite insensitive to the underlying
  structure of quantum phase space dynamics. We identify the flow that
  is the quantum analog of classical particle flow along phase
  portrait lines. It reveals hidden features of quantum dynamics and
  extra complexity. Being constrained by conserved flow winding
  numbers, it also reveals \new{fundamental topological order in 
    quantum dynamics that has so far gone unnoticed}.
\end{abstract}

 %Quantum mechanics 

\maketitle

Phase portraits characterise the time evolution of dynamical systems
and are widely used in classical mechanics~\cite{Nolte_PT10}.  For the
conservative motion of a single particle, moving in one
dimension~\nd{x} under the influence of a static smooth
potential~\nd{V(x)} only, the classical Liouville flow in phase space
is regular~\cite{Berry_AIPC78} and largely determined by location and
nature of its flow stagnation points. Situated on the $x$-axis
wherever the potential is force-free (momentum $p=0$ and $-\partial V
/ \partial x=0$), the local flow forms clockwise vortices around
stagnation points at minima of the potential, maxima split the flow
and therefore lie at the intersections of flow separatrices, saddle
points of the potential lead to an elongated saddle flow pattern
oriented along the $x$-axis.

Here we investigate the quantum dynamics of bound states of
nonharmonic potentials; their quantum phase space flow reveals rich
nonclassical features:
\\
Dependence of flow on the state of the system~\cite{Donoso_JCP03}
leading to directional deviation from classical
trajectories~\cite{Bauke_2011arXiv1101.2683B} and flow
reversal~\cite{footnote_harmonic_flow}, time-dependent
quantum displacement of classical stagnation
points~\cite{Skodje_PRA89}, occurrence of additional nonclassical
stagnation points (see Fig.~\ref{1} below) whose positions change over
time~\cite{Skodje_PRA89} (even for conservative systems), and
conservation of the flow orientation winding number $\omega$, see
Eq.~(\ref{eq:WindingNumber}) below, carried by all flow stagnation
points during all stages of their time evolution --including
instances when they split from or merge with other stagnation points.

For a single quantum particle described by a complex time-dependent
amplitude function~\nd{\psi(x;t)} the associated quantum analog of
classical phase space probability distributions is Wigner's
function~\nd{W(x,p;t)}~\cite{Wigner_32,Hillery_PR84}, with~\nd{p} the
particle's momentum. Structurally, $W$ is a Fourier transform of
the off-diagonal coherences of the quantum system's density
matrix~\nd{\varrho}, i.e.
\begin{equation}
  W(x,p;t) = \frac{1}{\pi \hbar} \int_{-\infty}^{\infty} dy \, 
  \varrho(x+y,x-y;t) \cdot e^{\frac{2i}{\hbar} p y} \; ,
\label{eq_WigFct_def}
\end{equation}
where \nd{\hbar=h/(2\pi)} is Planck's constant. Unlike $\psi$ or
$\varrho$, the Wigner function only assumes real values, but these do
become negative~\cite{Wigner_32,Schleich_01}, defying description in
terms of classical probability
theory~\cite{Schleich_01,Zurek_01,Grangier_SCI11}, thus revealing
quantum aspects of a system~\cite{Ferraro_PRL12}.

The time evolution of~$W$ can be cast in the form of a flow
field~\nd{\bm{J}(x,p;t)}~\cite{Donoso_PRL01}, the `Wigner
flow'~\cite{Bauke_2011arXiv1101.2683B}, \new{which describes the flow
  of Wigner's quasi-probability density in phase space. It} has the
two components
\begin{equation}
{\bm J} =  \left( J_x \atop J_p \right) = \left(\frac{p}{m} W(x,p;t) 
    \atop -\sum\limits_{l=0}^{\infty}{\frac{(i\hbar/2)^{2l}}{(2l+1)!}
      \frac{\partial^{2l}W(x,p;t)}{\partial p^{2l}}
      \frac{\partial^{2l+1}V(x)}{\partial x^{2l+1}}} \right) \; ,
\label{eq:FlowComponents}
\end{equation}
fulfilling Schr\"odinger's equation which takes the form
\new{
\begin{equation}
\frac{\partial W}{\partial t} + \frac{\partial
  J_x}{\partial x} + \frac{\partial J_p}{\partial p} =
0 \; 
\label{eq:Continuity}
\end{equation} 
of a continuity equation~\cite{Wigner_32}. Thus Wigner flow is the
equi\-valent of classical Liouville flow, it has, so far, not been
studied in great
detail~\cite{Donoso_PRL01,Hughes_JPC07,Gat_JPA07,Bauke_2011arXiv1101.2683B}.
\\
Nonlocality~\cite{Bell_Book87,Donoso_JCP03} originates both in
definition~(\ref{eq_WigFct_def}) of the Wigner function and the higher
derivatives of~$V$ occurring in the Wigner
flow~(\ref{eq:FlowComponents}).}

\new{The marginals of the Wigner function yield the probability
  distributions in position $|\psi(x;t)|^2$ and momentum
  $|\phi(p;t)|^2$, see Fig.~\ref{2} and
  references~\cite{Wigner_32,Schleich_01,Zachos_book_05}. Integrating
  over the expressions in the continuity equation analogously shows
  that the marginals of the Wigner flow yield the respective
  probability currents in $x$ and $p$
\begin{eqnarray}
 & \frac{d}{dt}|\psi(x;t)^2| = \int_{-\infty}^{\infty} dp \; J_x(x,p;t) = \hat \jmath_x(x;t) 
 \label{eq:ProbCurr_x} \\
 \mbox{and } \quad & 
 \frac{d}{dt}|\phi(p;t)^2| = \int_{-\infty}^{\infty} dx \; J_p(x,p;t) = \hat \jmath_p(p;t)  \; ,
\label{eq:ProbCurr_p} 
\end{eqnarray}
where $\phi(p)$ is the momentum representation of $\psi(x)$. While the
$x$-component can be rewritten as the probability current $ \hat
\jmath_x(x;t) = \frac{\hbar}{2im} \left( \psi^* \partial_x \psi -
  \psi \partial_x \psi^* \right) $, in general, no such simple
expression exists for $\hat \jmath_p(p;t)$.}

\begin{figure}[t!]
%  \centering
  {\includegraphics[width=\columnwidth]{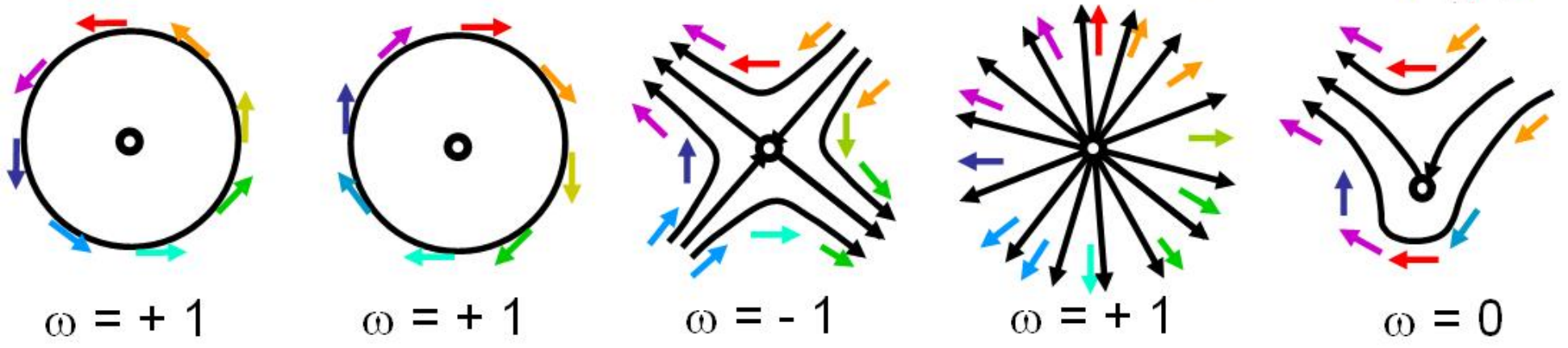}}
  \caption{(color online) Flow field around \new{various} types of
    stagnation points of Wigner flow with associated winding
    numbers. This list is nonexhaustive.}
\label{1}
\end{figure}

The dynamics of the harmonic potential, the most studied quantum case,
e.g. of quantum optics~\cite{Schleich_01,Grangier_SCI11}, amounts to a
rigid rotation of the Wigner function around the origin of phase
space. Only at the origin can a flow vortex form~(Fig.~3 in
Ref.\cite{Bauke_2011arXiv1101.2683B}), just like in the classical
case. The quantum harmonic oscillator and its
isomorphism~\cite{Ole_FreeHOSC_arxiv_10}, the free quantum
particle~\cite{Kurtsiefer_NAT97}, constitute exceptional, degenerate
cases where lines of stagnation of Wigner flow occur, and not only
isolated stagnation points.  This is due to the fact that for
$V\propto x^2$ or $V=$const. we have in Eq.~(\ref{eq:FlowComponents})
$ J_p = -W \frac{\partial V}{\partial x}$, just as in the `classical
limit'~($\hbar \rightarrow 0$) and, consequently, with $W=0$ we always
find $|\bm J| =0$. Because of these three facts~(rigid rotation,
classical form of $J_p$ and line formation) the nonclassical phase
space features discussed here cannot be seen in the degenerate cases
primarily studied so
far~\cite{Schleich_01,Kurtsiefer_NAT97,Grangier_SCI11}.
\begin{figure}[b!]
%  \centering
  {\includegraphics[width=\columnwidth]{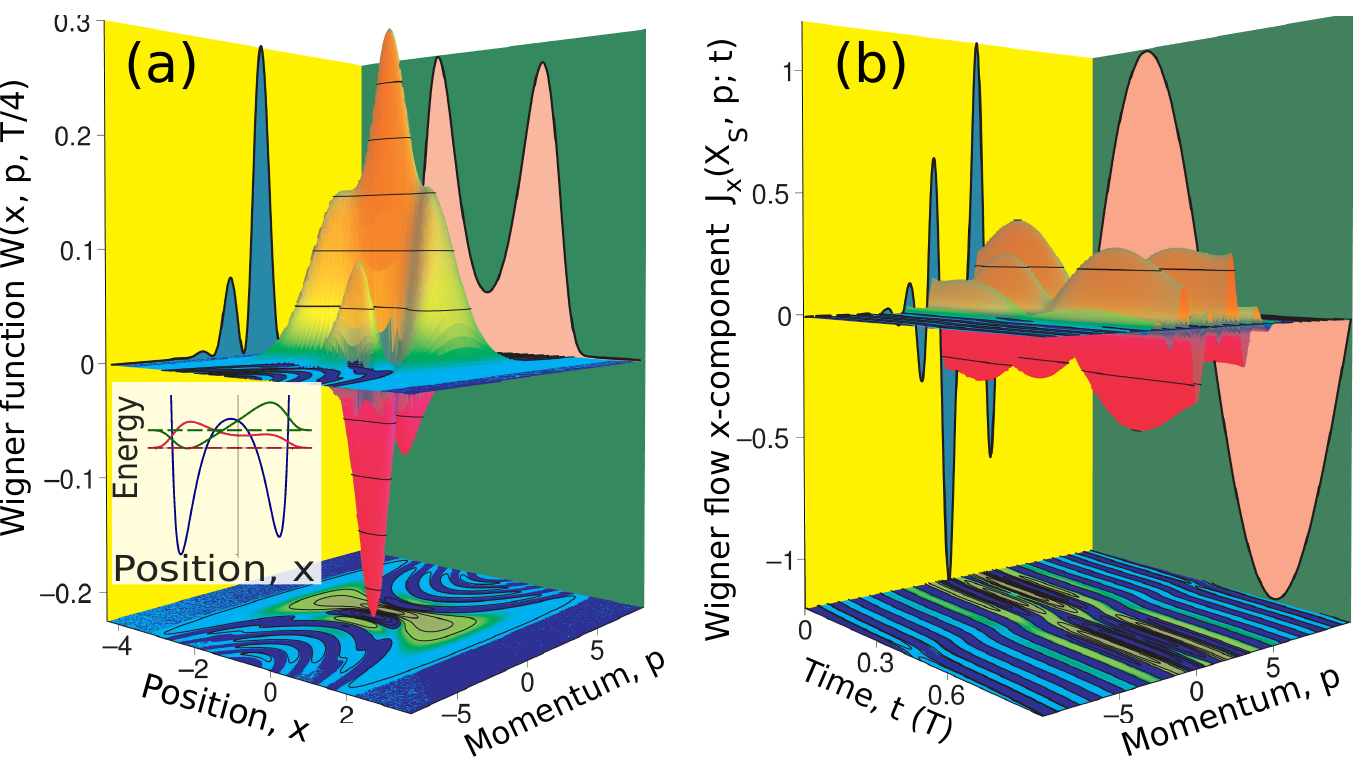}}
  \caption{(color online) {Wigner function, Wigner flow, and momentum
      distribution of state~$\Psi$ (parameters~$\hbar=1, m=1/2, \alpha
      = 0.5$, $\Delta E = 0.5$).} (a) Wigner function~$W(x,p;T/4)$,
    projection to bottom shows its contours. Projections onto
    background walls show momentum and position probability
    distribution (blue and rose filled curves, respectively, in
    arbitrary units). Inset: Plot of Caticha-potential with wave
    functions for lowest two energy eigenstates shifted to their
    respective energy levels (dashed lines). (b) $x$-component of
    Wigner flow~$J_x(X_S,p;t)$ at barrier top at position~$X_S$;
    bottom projection shows its contours (note its phase shift at
    $t=T/2$). Projections onto background walls show time and momentum
    projections (blue and rose filled curves, $\langle J_x(X_S,p)
    \rangle_T = \int_{0}^{T} d\tau \; J_x(X_S,p;\tau)$ \new{and
      $\hat\jmath_x(X_S;t)$ of Eq.~(\ref{eq:ProbCurr_x}),
      respectively, in arbitrary units). The
      projection~$\hat\jmath_x$} features the sinusoidal
    variation~$\propto \sin(2\pi t/T)$ expected of the tunnelling
    current of a two-state system. }
\label{2}
\end{figure}

The degeneracies of the degenerate cases leading to the formation of
stagnation lines are lifted for nonharmonic potentials by the
presence of terms with~$l>0$ in~$J_p$ and leads to the formation of
separate stagnation points instead. The boundedness and continuity of
wave functions of infinitely differentiable nonharmonic potentials
and the unitarity of such systems' quantum dynamics induces homotopies
that keep all smooth changes of~$\bm J$ in space and time around
stagnation points smooth. We therefore conjecture that the topological
structure of the Wigner flow field around stagnation points remains
conserved. To monitor this we introduce the Wigner flow orientation
winding number determined by the integral
\begin{eqnarray}
  \label{eq:WindingNumber}
  \omega({\cal L};t) =\frac{1}{2\pi} \varointctrclockwise_{\cal L} d
  \varphi \; 
\end{eqnarray} 
along a closed (convex) loop~$\cal L$; here~\nd{\varphi} is the
orientation angle between the positive $x$-axis and the Wigner flow
`vectors'~$\bm J$.  For `empty' paths, not including a stagnation
point of~$\bm J$, $\omega=0$; for vortices $\omega = +1$, see
Fig~\figref{2}. The winding number~$\omega({\cal L};t)$ is unchanged
under continuous path deformations that do not push~${\cal L}$ across
a stagnation point and as long as time evolution does not move a
stagnation point across the loop. It assumes integer values only
(assuming the integration path~${\cal L}$ does not run through a
stagnation point) and is conserved. The sum of winding numbers of all
stagnation points within a loop is conserved, even when they split or
coalesce: the stagnation points carry topological
charge~\cite{Dennis_PO09}.

\begin{figure}[t!]
  {\includegraphics[width=\columnwidth,height=0.9\columnwidth]{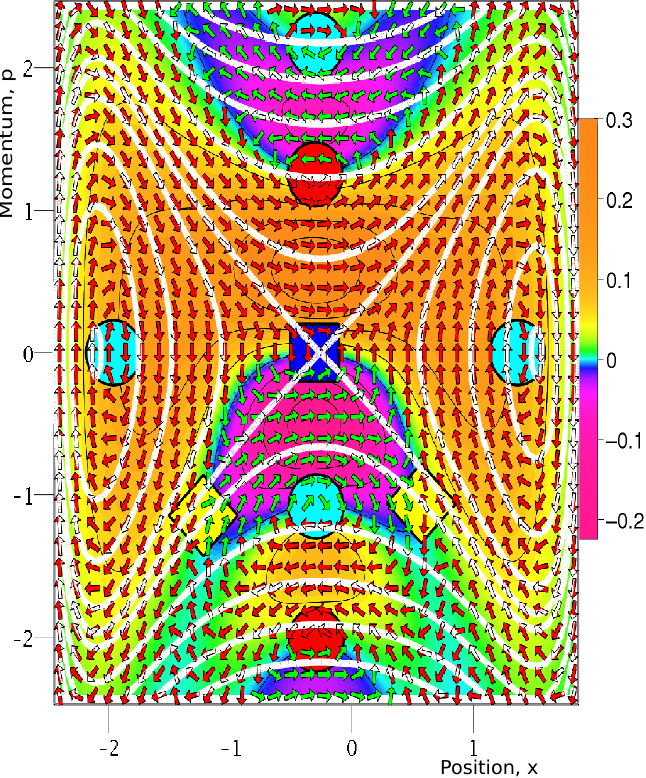}}
  \caption{(color online) {Features of Wigner flow of an asymmetric
      double well potential (same parameters as in Fig.~1, at
      $t=T/4$). Contour plot of Wigner function~$W(x,p;T/4)$ (black
      contour lines) with an overlay of coloured arrows showing
      normalised Wigner flow $\bm J/|\bm J|$ (red arrows for $W>0$ and
      green for $W<0$).}  The classical phase portrait is shown as a
    collection of thick white lines. All locations of Wigner flow
    stagnation points are highlighted by symbols (cyan and red circles
    centre on clockwise and anticlockwise vortices, respectively,
    yellow diamonds on sepa\-ra\-trix intersections, and the blue
    square on a $p$-directed saddle flow. The quantum displacement of
    the vortices near the potential minima, towards the center, is
    clearly visible.}
\label{3}
\end{figure}

To give an example, we concentrate on Caticha's~\cite{Caticha_95}
smooth, slightly asymmetric, double well potential
\begin{eqnarray}
  \label{eq:AsymmetricPotential}
\nonumber  V(x) & = & 1 +  E_0 + \frac{3}{2}\Delta E - \Delta E\alpha\sinh (2  x)
\\ \nonumber
  & + &\cosh^2 (  x)\left(\frac{\Delta E^2}{4 }\alpha\sinh
    (2  x) - \frac{\Delta E^2}{4}  - 2\Delta E\right) 
\\ %\nonumber
& + & 
 \frac{\Delta E^2}{4 }\left( \alpha^2 + 1\right)\cosh^4
  ( x)%\\
 \; ,
\end{eqnarray}
featuring high outer walls, and wells separated by a barrier of
sufficient height, such that at least ground and first excited energy
eigenstate tunnel through it,~Fig.~\figref{2}(a)~inset. To display all
nonclassical flow features listed in the introduction it suffices to investigate
the balanced superposition
\begin{eqnarray}
  \label{eq:Balanced_Psi}
\Psi(x;t)=  \frac{\psi_0(x) e^{-iE_0t/\hbar} - \psi_1(x) e^{-i(E_0+\Delta
E)t/\hbar}}{\sqrt 2}\; ,
\end{eqnarray}
of ground~$ \psi_0(x) = \psi_0 \cosh (x) \exp\left[ -\frac{\Delta E}{4
  } \left(\cosh^2 ( x) + \alpha x \right. \right.
\\
\left. \left. + \frac{\alpha}{2}\sinh (2 x)\right)\right] $ and first
excited state~$ \psi_1(x) = \psi_1 \left[ \alpha + \tanh(x) \right]
\psi_0 (x)$, with energies~\nd{E_0} and~\nd{E_0+\Delta E} and
normalisation constants~\nd{\psi_0} and~\nd{\psi_1},
respectively~\cite{Caticha_95}. \new{Since the eigenstates are real functions their Wigner functions obey $W(x,p) = W(x,-p)$ which implies that
\begin{eqnarray}
 & J_x(x,p) = - J_x(x,-p) \label{eq:FlowComponentSymmetries_x}\\
\mbox{and } &  J_p(x,p) =  J_p(x,-p)  \; .
\label{eq:FlowComponentSymmetries_p}
\end{eqnarray}
Their individual Wigner flow patterns, which the superposition~$\Psi$
`inherits', are displayed in the
supplement~\cite{supplement_eigenstates}. All these} Wigner functions
have to be determined numerically~\cite{numerics}. Of the low energy
states~$\Psi$ is the `most dynamic' in that it shifts all of the
particle's probability distribution back and forth between left and
right well. Its Wigner function $W(t=T/4)$ at a quarter of the
tunnelling period time ($T=2\pi\hbar/\Delta E$) is displayed in
Figs.~\figref{2}(a) and~\figref{3}.

The Wigner flow's state dependence has two aspects: the negativity of
$W$ reverses the flow, and the dependence of $J_p$ on $W$ and $V$
leads to sideways deviation of Wigner flow from classical phase
portrait lines. In the case of eigenstates of the harmonic oscillator,
the former leads to shear flow between neighbouring sectors of
alternating polarity~\cite{Bauke_2011arXiv1101.2683B}. The latter
deviation can remain mild for eigenstates of a weakly anharmonic
potential~\cite{Bauke_2011arXiv1101.2683B}, in our case it is very
pronounced leading to the formation of nonclassical vortices which are
quantum displaced off the $x$-axis and some of which spin
anticlockwise,~Figs.~\figref{1} and~\figref{3}. Indications of
nonclassical vortices seem to have been observed in `chaotic' quantum
systems before~\cite{Berry_JPA79}.

\begin{figure}[t!]
  %\centering
  {\includegraphics[width=\columnwidth]{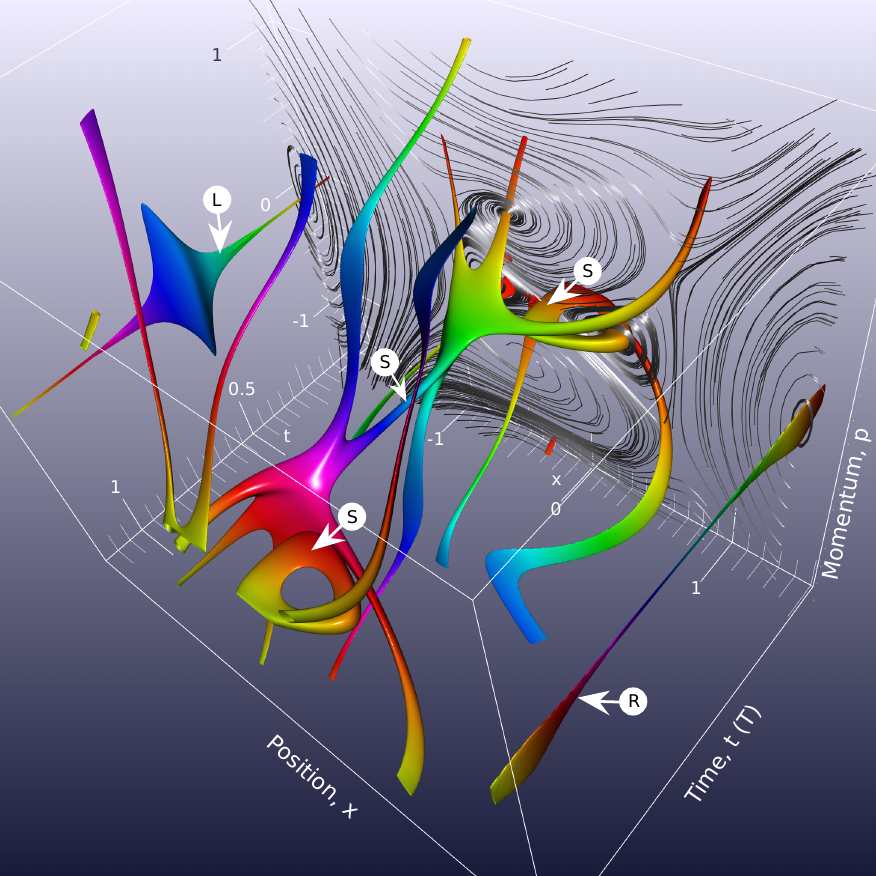}}
  \caption{(color online) {Wigner flow's stagnation points' positions
      across phase space as a function of time.} Same parameters as in
    Fig.~\ref{2}. The tube surfaces show where in phase space the
    magnitude of Wigner flow is small~${\bm J}(x,p;t)^2=3 \times
    10^{-5}$. They are displayed over 120\% of one oscillation period
    ($t=-0.1 T \ldots 1.1 T$); the rainbow spectrum is matched to $T$,
    red-orange for $t=0$, via green, cyan at $t=0.5 T$, through blue
    and purple back to red-orange. Because of the periodicity of the
    two-state scenario, the red-orange-yellow torus is seen twice at
    beginning and end of the time window.  At the core of all tubes
    lie time lines of stagnation points. Their movement through phase
    space leads to their mergers and splits. The grey flow integration
    lines %~\cite{Benger_VISH07}
    are guiding the eye past vortices and separatrices, they do not
    represent physical flow since they integrate~${\bm J}(x,p;t=0.05)$
    at fixed time. The $x$-coordinate is shown from left well minimum
    at~$X_L=-2.095$ to right well minimum at~$X_R=1.514$, the position
    of the associated vortices ($\bigcirc \! \!  \!\! \!  \mbox{\small
      L}$ in left well and $\circledR$ in right), at $p=0$ and just
    inside the plot region, confirms the inward quantum displacement
    of these stagnation points. The remnant of the classical
    separatrix stagnation point at position $X_S+\delta x_S(t) \approx
    -0.3$ is labelled $\circledS$. It follows a bent path and becomes
    displaced, forming part of the torus, when it coalesces with or
    splits from other stagnation points. All mergers or splits are
    constrained by topological charge~\cite{Dennis_PO09} conservation,
    see Fig.~\figref{5}.}%
\label{4}
\end{figure}

Flow reversal affects quantum tunnelling. The wave function of a
particle, tunnelling through a barrier, is coherent across it,
implying that interference fringes of the Wigner function form in the
tunnelling region; orientated parallel to the
$x$-axis~\cite{Zurek_01},~Figs.~\figref{2}(b)
and~\figref{3}. Neighbouring phase space regions contain strips of
alternating Wigner function polarity alternating their flow
direction~\cite{Skodje_PRA89},~Fig.~\figref{2}(b). The further apart
the two wells, the finer the interference pattern~\cite{Zurek_01} and
the stronger the resul\-ting flow cancellation.  Thus we find that
tunnelling can be described as a transport phenomenon that plays out
over large parts of quantum phase
space~\cite{Balazs_AP90,Skodje_PRA89}, frustrated by phase space
interference.
 
%\onecolumngrid
%\begin{figure}[t!]
\begin{figure*}[t!]
%    {\includegraphics[width=18cm,height=11cm]{Fig_5}}
    {\includegraphics[width=17.5cm]{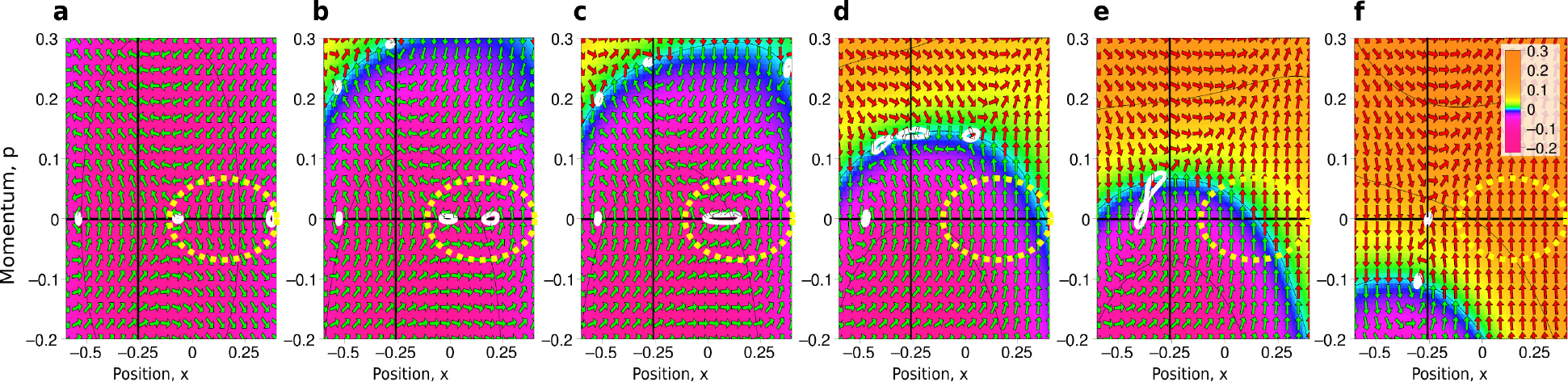}}
%    {\includegraphics[width=12cm,height=5.7cm]{Fig_5}}
    \caption{(color online) {Same system parameters as for
        Fig.~\ref{2}: Contour plots of Wigner function with
        superimposed normalised flow field~ $\bm J/ |\bm J|$ at times
        $t = 0, 0.075T, 0.0875T, 0.15T, 0.2T,$ and $0.325T$ in panels
        {\bf a} to {\bf f} demonstrates mergers and splitting of
        stagnation points and their pinning to zeros of the Wigner
        function.} Each vertical black line shows the position~$X_S$
      of the tunnelling barrier top. Small white loops delineate areas
      where {$\bm J^2 = 2 \times 10^{-6}$} and encircle stagnation
      points. We chose the region where the torus closes and several
      stagnation points coalesce (compare Fig.~\figref{4}). The cross
      sections of the torus are visible in the right half of panels
      {\bf a}, {\bf b} and~{\bf c}, a convex loop~$\cal L$ encircling
      both torus points only, such as the dotted yellow ellipse,
      yields~$\omega=0$. This value is conserved over time throughout
      the merger which leads to the disappearance of the torus
      (panel~{\bf d}). The mergers of the four stagnation points
      outside the yellow loop (two vortices and two separatrix
      crossings) in panels {\bf d}, {\bf e} and {\bf f} into one
      vortex and one separatrix crossing evidently carry a conserved
      total topological charge of zero throughout.}%
\label{5}
\end{figure*}
%\twocolumngrid
%
\noindent

For a low energy state, such as~$\Psi$, the positions of remnant
vortices originating from the potential's minima positions are
quantum displaced inward, towards the potential barrier,
Figs.~\figref{3} and~\figref{4}. 

In classical physics stagnation points of phase space flow can only
occur on the $x$-axis, here, according to
Eq.~(\ref{eq:FlowComponents}), they are pinned to the zero lines of
the Wigner function, but occur off the $x$-axis, travel long distances
and merge with or split from other stagnation points, see
Figs.~\figref{4} and~\figref{5}.

For the Caticha potential, we observe a string of vortices with
alternating handedness aligned in the $p$-direction located near the
top of the tunnelling barrier, at
position~$X_S=-0.258$,~Fig.~\figref{3}. Over time they travel in the
negative $p$-direction, Figs.~\figref{4} and~\figref{5}. When reaching
the $x$-axis they coalesce with the remnant separatrix intersection
point while the overall topological charge of the flow is conserved;
see Fig.~\figref{5}.

To conclude, Wigner flow reveals nonclassical features and added
complexity of quantum phase space dynamics. At the same time it
provides, through the conservation of the flow winding number
$\omega$, the basis for an analysis of its topological invariants,
ordering this complexity.

Systems that have been studied using quantum phase space techniques
can be analysed using Wigner flow. Such systems
arise~\cite{Zachos_book_05} for example, in chemical quantum
dynamics~\cite{Donoso_PRL01,Hughes_JPC07,Gibson_JCP86}, ``nonlinear''
quantum processes in closed single particle~\cite{Gat_JPA07} or open
multiparticle~\cite{Garraway_PRA94,Katz_NJP08,Rips_NJP12} systems,
classical electromagnetic fields~\cite{Levanda_AP01} and multiband
semiconductor physics~\cite{Morandi_PRB09}.

\begin{acknowledgments}
  Use of the high-performance computing facility at the University of
  Hertfordshire's Science and Technology Research Institute is
  gratefully acknowledged. We thank Charles Young, Elias Brinks,
  Daniel Polani, Andreas Kukol and Cosmas Zachos for comments on the
  manuscript.
\end{acknowledgments}

\bibliography{master}

\begin{thebibliography}{31}
\bibitem [{foo()}]{footnote_harmonic_flow}%
  \BibitemOpen
  \href@noop {} {}\bibinfo {note} {{In degenerate quantum cases flow reversal
  can occur but none of the other nonclassical flow patterns listed
  here.}}\BibitemShut {Stop}%
\bibitem [{num()}]{numerics}%
  \BibitemOpen
  \href@noop {} {}\bibinfo {note} {{To determine the Wigner functions we
  integrated $y$ from -2.7$,\ldots,2.7$. The summation cutoff for the
  determination of $J_p$ was chosen at $l_{max}=12$.}}\BibitemShut {Stop}%
\end{thebibliography}

\cleardoublepage
\onecolumngrid
\pagenumbering{gobble}

\section*{{\sf Supplement to:} \\ 
\vspace{1cm}
{\bf 
Wigner flow reveals topological order in quantum phase space dynamics}
 \\
\vspace{0.3cm}
{\sf by Ole Steuernagel, Dimitris Kakofengitis \& Georg Ritter}
}

In this supplement we display the Wigner flow patterns of the ground
state~$\psi_0$ and first exited state~$\psi_1$ constituting the
superposition state~$\Psi $ of equation~(\ref{eq:Balanced_Psi}).

\setcounter{section}{1}
\renewcommand{\thesection}{S\arabic{section}}
\renewcommand{\thefigure}{\thesection}
\renewcommand{\thefigure}{S}

\begin{figure}[h!]
  \centering
  {\includegraphics[width=15cm]{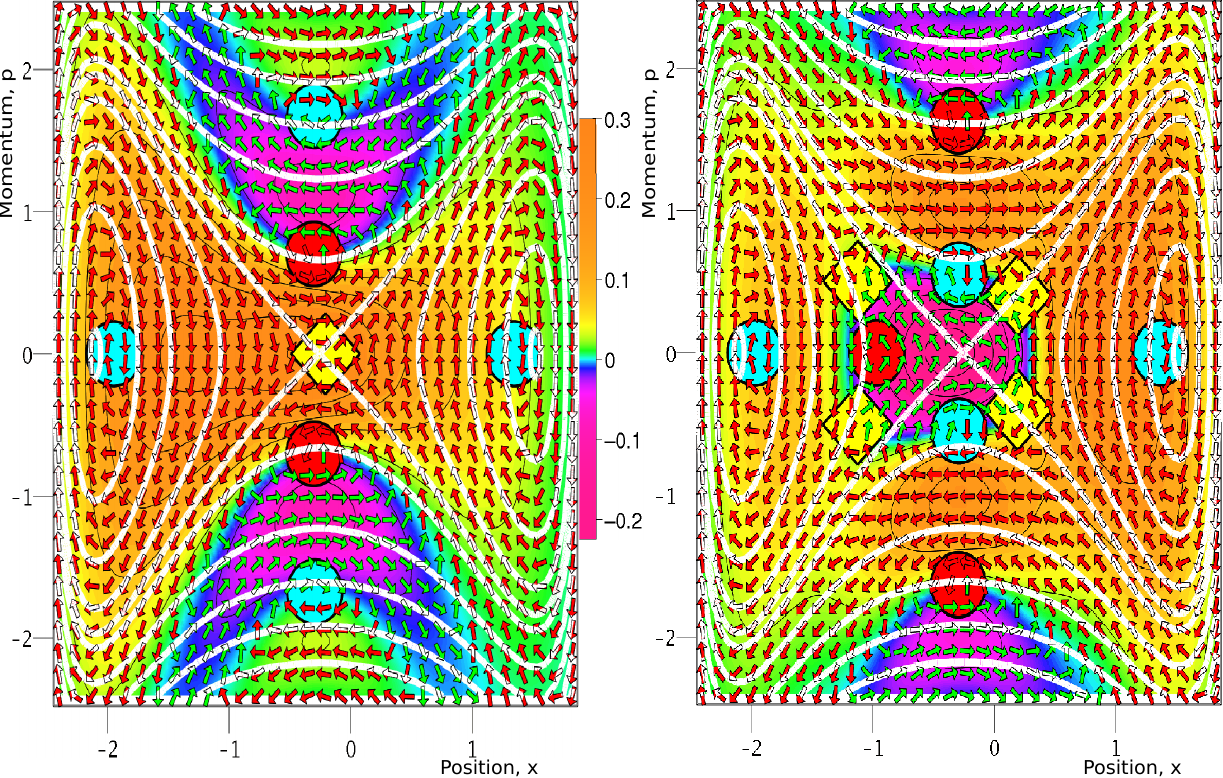}}
  \caption{The Wigner flow of the eigenstates~$\psi_0$ and~$\psi_1$
    displays the symmetries of
    equations~(\ref{eq:FlowComponentSymmetries_x})
    and~(\ref{eq:FlowComponentSymmetries_p}). The same parameters and
    color coding as in Fig.~\figref{3} are used.}
\label{C}
\end{figure}

\end{document}